\documentclass[twocolumn,showpacs,preprintnumbers,superscriptaddress,amsmath,floatfix,amssymb,prd]{revtex4}
\usepackage{graphicx}
\usepackage[colorlinks=true]{hyperref}
\usepackage{amsfonts,amsmath,amsxtra}

\newcommand{\beq}[1]{\begin{eqnarray}\label{#1}}
\newcommand\eeq {\end{eqnarray}}
\newcommand\bqa {\begin{eqnarray}}
\newcommand\eqa {\end{eqnarray}}

\newcommand{\bear}{\begin{array}}
\newcommand{\enar}{\end{array}}

\newcommand{\R}{\mathbb{R}}
\newcommand{\C}{\mathbb{C}}
\newcommand{\K}{\mathcal{K}}

\newcommand{\Z}{\mathbb{Z}}

\newcommand{\N}{\mathbb{N}}

\begin{document}
\def\le{\langle}
\def\re{\rangle}
\def\dg{^{\dag}}
\def\K{{\cal K}}
\def\n{{\cal N}}
\font\maj=cmcsc10 \font\itdix=cmti10
\def\N{\mbox{I\hspace{-.15em}N}}
\def\H{\mbox{I\hspace{-.15em}H\hspace{-.15em}I}}
\def\1{\mbox{I\hspace{-.15em}1}}
\def\pN{\rm{I\hspace{-.1em}N}}
\def\Z{\mbox{Z\hspace{-.3em}Z}}
\def\pZ{{\rm Z\hspace{-.2em}Z}}
\def\R{{\rm I\hspace{-.15em}R}}
\def\pR{{\rm I\hspace{-.10em}R}}
\def\C{\hspace{3pt}{\rm l\hspace{-.47em}C}}
\def\Q{\mbox{l\hspace{-.47em}Q}}
\def\b{\begin{equation}}
\def\e{\end{equation}}
\def\bee{\begin{enumerate}}
\def\wt{\widetilde}

\title{Higher Order Corrections to Asymptotic-de Sitter Inflation}
\author{M. Mohsenzadeh}
\email{mohsenzadeh@qom-iau.ac.ir} \affiliation{Department of Physics, Qom Branch, Islamic Azad University, Qom, Iran}
\author{E. Yusofi}
\email{e.yusofi@iauamol.ac.ir} \affiliation{Department of Physics,
Ayatollah Amoli Branch, Islamic Azad University, Amol, Mazandaran,
Iran}
\affiliation{School of Astronomy, Institute for Research in Fundamental Sciences(IPM), P. O. Box 19395-5531,Tehran, Iran}
\date{\today}

\begin{abstract}
Since  trans-Planckian considerations can be associated with the re-definition of the initial vacuum, we investigate further the influence of trans-Planckian physics on the spectra produced by the initial quasi-de Sitter (dS) state during inflation. We use the asymptotic-dS mode to study the trans-Planckian correction of the power spectrum to the quasi-dS inflation. The obtained spectra consist of higher order corrections associated with the type of geometry and harmonic terms sensitive to the fluctuations of space-time (or gravitational waves) during inflation. As an important result, the amplitude of the power spectrum is dependent on the choice of $c$, i.e. the type of space-time in the period of inflation. Also, the results are always valid for any asymptotic dS space-time and particularly coincide with the conventional results for dS and flat space-time.

\end{abstract}
Submitted to Chinese Physics C


\maketitle
\section{Introduction}

Recently, it has been realized that much can be learned about high energies and small scales by studying the very early universe \cite{c1, c2, c3}. Nevertheless, the physics of the very early universe is described by inflationary scenarios \cite{c4, c5, c6, c7}. Inflation is postulated to have caused small quantum fluctuations in the inflaton field to be magnified, consequently creating CMB anisotropies and large structures in the universe \cite{c8, c9}. The interesting scales in the recent universe correspond to the very short perturbations which fell under the Planck length at the start of inflation. However, in most inflationary models, the adiabatic vacuum \cite{c1, c10} is chosen as the quantum state of the inflation and the Bunch-Davies initial mode is imposed at the moment which the scales of most perturbations are smaller than the Planck length \cite{c2, c11}. General relativity and the theory of linear perturbations which are the backbone of the inflationary models, however, are collapsed at trans-Planckian scales. Indeed, trans-Planckian scales may be explained by incomplete theories such as quantum gravity, non-commutative geometry of space-time and string theory \cite{c12, c13, c14}. So, one may ask, ``how reliable are the predictions of the ordinary inflationary models?'' Danielsson answered this question by re-investigating the vacuum definition to include the trans-Planckian corrections \cite{c1}. Indeed, in the Danielsson approach, the initial condition is imposed at the moment at which the majority of perturbations are inside the horizon but their scales are greater than the Planck length~\cite{c1}. It is clear that the new vacuum, the so-called $ \alpha$-vacuum, results in an oscillatory power spectrum which is no  longer scale-invariant anymore~\cite{c1}. In the definition of the $ \alpha$-vacuum, it is supposed that the geometry of the background space-time is described by a pure de Sitter (dS) metric. It is important to note that the deviation of the initial condition for primordial perturbations from the Bunch-Davies vacuum were often due to some dynamical reason. In particular, the universe may have experienced other stages prior to the inflationary phase, namely, a fast-roll pre-inflationary phase \cite{c15}; a phase transition from fundamental theory \cite{c16, c17}; a modulation of the sound speed parameters \cite{c18}; or a nonsingular bouncing phase \cite{c19, c20}. Also, some alternative non-trivial initial states have been studied by many authors \cite{c21, c22,c23,c24,c25,c26,c27,c28,c29,c30,c31,c32,c33}.\\
 According to the slow-roll model, the geometry of the inflationary universe is described by quasi-dS space-time, which when considering only the first order reduces to pure dS space-time, so the $ \alpha$-vacuum should be modified for use in real inflationary models \cite{c34}.  However, due to the high energy level of the inflationary universe, higher order perturbations seem perfectly plausible, so we proposed a second order vacuum which may explain the origin of non-Gaussianity and scale dependency of the primordial perturbations \cite{c35, c36}. This initial mode so-called excited-dS vacuum is the generalization of the Bunch-Davies vacuum to the quasi-dS inflation. The scalar power spectrum deduced from this initial mode is not scale-invariant~\cite{c35, c36, c37, c38, c39}. Meanwhile, against the $ \alpha$-vacuum case, the calculated spectra included higher order terms of corrections. Something like this spectrum has been derived previously in \cite{c1, c2}. Furthermore, similar corrections have been obtained in Refs.~\cite{c40, c41} with an effective field theory approach.\\
Since the excited mode that we introduced in Ref.~\cite{c39} is an approximate solution of the Mukhanov-Sasaki equation, in this work we plan to use the general initial modes with higher order term of $\frac{1}{k\tau}$ to drive trans-Planckian corrections in the spectra. These forms of initial modes, as the general solution of the Mukhanov-Sasaki equation in quasi-dS inflation, are obtained from asymptotic expansion of the Hankel function \cite{c37, c42}. The layout of the paper is as follows. In Section 2, we briefly recall the definition of the standard power spectrum and review the Danielsson approach as well as the vacuum concept. Section 3 is the main part of this article, containing the calculation of the power spectrum with an asymptotic higher order initial mode. The conclusion is given in the final section.

\section{Review of Danielsson Approach for $ \alpha$-vacuum}
Suppose that the geometry of the inflationary universe is described by the following dS metric:
\begin{equation} ds^{2}=dt^{2}-e^{2Ht}{\vec{dx}}^{2}.\end{equation}
Here $ t $ and $ H $ are respectively time and the Hubble parameter during inflation. The matter field of the universe in the inflationary epoch may be explained by a massless single scalar field $ \phi(t,\mathbf{x}) $ with Lagrangian density
\begin{equation} \ell=-\frac{\sqrt{-g}}{2}g^{\alpha\beta}\partial_{\alpha}\phi \partial_{\beta}\phi.\end{equation}
Here we suppose $ \phi $ is minimally coupled to gravity, so
\begin{equation} {\phi}''+2\frac{a'}{a}{\phi}'-\nabla^{2}\phi=0, \end{equation}
where the prime symbol stands for the derivative with respect to the conformal time $ \tau=-\frac{1}{aH}$. Equation (3) in terms of the rescaled scalar field $ \mu=a\phi $ reduces to
\begin{equation} {\mu}''-\nabla^{2}\mu-\frac{a''}{a}{\mu}=0, \end{equation}
or in Fourier space
\begin{equation} {\mu''_{k}}+(k^{2}-\frac{a''}{a}){\mu_{k}}=0. \end{equation}
It is easy to show that $ \Pi_{k}={\mu'_{k}}-\frac{a'}{a}{\mu_{k}} $ is the conjugate momentum of $ \mu_{k} $, so we can quantize  similar to the simple harmonic oscillator\cite{c1},
\begin{equation} \hat{\mu_{k}}(\tau)=\frac{1}{\sqrt{2k}}[\hat{a}_{k}(\tau)+\hat{a}_{-k}^{\dag}(\tau)], \end{equation}
\begin{equation} \hat{\Pi_{k}}(\tau)=-i\sqrt{\frac{k}{2}}[\hat{a}_{k}(\tau)+\hat{a}_{-k}^{\dag}(\tau)], \end{equation}
where $ \hat{a}_{k}(\tau) $ and $ \hat{a}_{-k}^{\dag}(\tau) $ are the annihilation and creation operators. Furthermore, according to the Bogoliubov transformations we have
\begin{equation} \hat{a}_{k}(\tau)=u_{k}(\tau) \hat{a}_{k}(\tau_{0})+v_{k}(\tau) \hat{a}_{-k}^{\dag}(\tau_{0}),\end{equation}
\begin{equation} \hat{a}_{-k}^{\dag}(\tau)=u_{k}^{*}(\tau) \hat{a}_{-k}^{\dag}(\tau_{0})+v_{k}^{*}(\tau) \hat{a}_{k}(\tau_{0}).\end{equation}
Here $ \tau$ is conformal time and $ \tau_{0} $ is an arbitrary fixed time. So, we can find
\begin{equation} \hat{\mu}_{k}(\tau)=f_{k}(\tau) \hat{a}_{k}(\tau_{0})+f_{k}^{*}(\tau) \hat{a}_{-k}^{\dag}(\tau_{0}),\end{equation}
\begin{equation} \hat{\Pi}_{k}(\tau)=-i[g_{k}(\tau) \hat{a}_{k}(\tau_{0})-g_{k}^{*}(\tau) \hat{a}_{-k}^{\dag}(\tau_{0})],\end{equation}
where
\begin{equation} f_{k}(\tau)=\frac{1}{\sqrt{2k}}[u_{k}(\tau)+v_{k}^{*}(\tau)], \end{equation}
\begin{equation} g_{k}(\tau)=\sqrt{\frac{k}{2}}[u_{k}(\tau)-v_{k}^{*}(\tau)]. \end{equation}
Notice that $ f_{k}(\tau_{0}) $ is a solution of Equation (5). Furthermore, from (8) and (9) one can find

$$ v_{k}(\tau_{0})=v_{k}^{*}(\tau_{0})=0, $$
and
$$ |u_{k}(\tau)|^{2}-|v_{k}(\tau)|^{2}=1, $$
or equivalently,
\begin{equation} g_{k}(\tau_{0})= k f_{k}(\tau_{0}),\end{equation}
\begin{equation} 2Re (f_{k}g_{k}^{*})=1. \end{equation}
Now let us define the initial mode as
\begin{equation} \hat{a}_{k}(\tau_{0})| 0,\tau_{0} \rangle=0. \end{equation}
This definition leads to a class of vacuums for the dS space-time which depend on initial fixed time $ \tau_{0} $. The initial mode defined in (16) is known as the minimum uncertainty vacuum, since it can be shown that it minimizes the uncertainty at $ \tau_{0} $~\cite{c1}.\\
For pure dS space-time, we have the following equation of motion
 \begin{equation} \label{Spa361}{\mu''_{\mathbf{k}}}+(k^{2}-\frac{2}{\tau^{2}}){\mu_{\mathbf{k}}}=0, \end{equation}
 and the most general solution of (\ref{Spa361}) is~\cite{c1}
\begin{equation} f_{k}=\frac{A_{k}}{\sqrt{2k}}e^{-{ik\tau}}(1-\frac{i}{k\tau})+\frac{B_{k}}{\sqrt{2k}}e^{{ik\tau}}(1+\frac{i}{k\tau}),\end{equation}
\begin{equation} g_{k}=A_{k}\sqrt{\frac{k}{2}}e^{-{ik\tau}}-B_{k}\sqrt{\frac{k}{2}}e^{{ik\tau}}.\end{equation}
Using normalization equation (15) as well as initial condition (14), one can find
\begin{equation} |A_{k}|^{2}=\frac{1}{1-|\alpha_{k}|^{2}}\;  , \;\; \alpha_{k}=\frac{i}{2k\tau_{0}+i}\;,\end{equation}
\begin{equation} B_{k}=A_{k}\alpha_{k}e^{-2ik\tau_{0}}. \end{equation}
If one sets $ \tau_{0}=-\infty $, this results in $ A_{k}=1 $ and $ B_{k}=0 $, and the Bunch-Davies vacuum is derived again. However,  instead of $ \tau_{0}=-\infty $, Danielsson supposed that $ \tau_{0}=-\frac{\Lambda}{kH} $ where $ \Lambda $ is the energy scale of new physics e.g. the Planck scale or the string scale. Every $ \tau_{0} $ defines a new initial mode and $ -\frac{\Lambda}{kH} $ defines the  $ \alpha $-vacuum~\cite{c1}. Next we can calculate the spectrum of $ \phi $,
\begin{equation} P_{\phi}=\frac{1}{a^{2}}P_{\mu=a\phi}=\frac{k^{3}}{2\pi^{2}a^{2}}|f_{k}|^{2}. \end{equation}
By considering those terms which are leading at late times when $ \tau_{0}\rightarrow 0,$  we have
\begin{equation} P_{\phi}=(\frac{H}{2\pi})^{2}\frac{1}{1-|\alpha_{k}|^{2}}[1+|\alpha_{k}|^{2}-e^{2ik\tau_{0}}\alpha_{k}^{*}-e^{-2ik\tau_{0}}\alpha_{k}] .\end{equation}
It is easy to see that by selecting $ \alpha_{k}=0 $ the spectrum reduces to
\begin{equation} P_{\phi}=(\frac{H}{2\pi})^{2}, \end{equation}
which coincides with the standard scale-invariant spectrum. If we assume $ -k\tau_{0}=\frac{\Lambda}{H_{0}}=\frac{1}{\sigma_{0}}\gg 1 $, as considered in Refs.~\cite{c1, c2}, we get
\begin{equation} y_1=(\frac{2\pi}{H})^{2}P_{\phi}=[1-\sigma_{0}sin(\frac{2}{\sigma_{0}})], \end{equation}
which is the Danielsson spectrum~\cite{c1}. The value of $\sigma_{0}=\frac{H_{0}}{\Lambda}$ is dimensionless, where $H_{0}$ is the initial Hubble expansion rate during inflation, and $\Lambda$ is the new physics energy scale. Spectrum (25) is vividly oscillatory due to the trans-Planckian corrections considered (notice that $ k=-\frac{\Lambda}{H_{0}\tau_{0}} $ ), so the wavelength of perturbations is proportional to $ \sigma_{0}\tau_{0}$,  which is greater than the Planck length.

\section{Asymptotic-dS Vacuum for Quasi-dS Inflation}
 Inflation predicts that the CMB temperature fluctuations should be: (i) statistically isotropic, (ii) Gaussian, and (iii) almost scale invariant \cite{c43}, and that this scale dependency can be describe theoretically by slow roll or quasi-dS inflation. If we suppose that cosmic inflation is described by dS space-time, there exists a concrete set of vacuum states invariant under the dS group. However, as we know, an inflationary universe is described by quasi-dS space-time, that may be exact dS in the first approximation \cite{c37}. Recently, as an approximate solution for quasi-dS space-time, we have added a perturbative second order term $\frac{1}{2}(\frac{\pm{i}}{k\tau})^{2}$ to the dS mode function (18) in \cite{c38} and have derived non-linear trans-Planckian corrections in the spectra.\\
Let us consider the general form of the Mukhanov-Sasaki equation for quasi-dS inflation \cite{c29} as the following form,
\begin{equation} \label{Muk25}  \upsilon''_{k}+(k^{2}-\frac{2c}{\tau^2})\upsilon_{k}=0, \end{equation}
where $c$ in terms of the Henkel function index $(\nu)$ can be written as follows \cite{c10},
 \begin{equation}
 \label{alf26}  c =\frac{4\nu^2-{1}}{8}.
 \end{equation}
  The general solutions of the mode equation (\ref{Muk25}) can be written as \cite{c34}:
  \begin{equation}
\label{Han22}
\upsilon_{k}=\frac{\sqrt{\pi \tau}}{2}\Big(A_{k}H_{\nu}^{(1)}(|k\tau|)+B_{k}H_{\nu}^{(2)}(|k\tau|)\Big), \end{equation} where
$ H_{\nu}^{(1, 2)} $ are the Hankel functions of the first and second kind, respectively.
Let us consider the general form of the mode function by expanding the Hankel functions up to the higher order of ${1}/{|k\tau|}$,
  $$
  \upsilon^{gen}_{k}(\tau, \nu)=A_{k}\frac{e^{-{i}k\tau}}{\sqrt{2k}}\big(1-i\frac{c}{k\tau}-\frac{d}{k^2\tau^2}-...\big)$$
 \begin{equation} \label{gen27} + B_{k}\frac{e^{{i}k\tau}}{\sqrt{2k}}\big(1+i\frac{c}{k\tau}-\frac{d}{k^2\tau^2}+...\big).
 \end{equation}
Note that $d ={c(c-1)}/{2}$. If we consider the special case of the pure dS space-time ($\nu={3}/{2}$ or $c =1$), the general form of the mode functions (\ref{gen27}) leads to the exact dS mode or $\alpha$-vacuum:
\begin{equation} \label{gen271}  \upsilon^{dS}_{k}(\tau, \nu)=A_{k}\frac{e^{-{i}k\tau}}{\sqrt{2k}}\big(1-i\frac{1}{k\tau}\big)+ B_{k}\frac{e^{{i}k\tau}}{\sqrt{2k}}\big(1+i\frac{1}{k\tau}\big).
 \end{equation}
Note that the general modes $\upsilon^{gen}_k$, are unable to recover the excited mode $\upsilon_k$ that was proposed in our previous work~\cite{c38}.
 The positive frequency solutions of the dS mode lead to the Bunch-Davies mode:
  \begin{equation}
 \label{Bun29} \upsilon_{k}^{BD}=\frac{1}{\sqrt{2k}}(1-\frac{i}{k\tau})e^{-ik\tau}.
\end{equation}
For this case, one has $a(t)=e^{Ht}$, or $ a(\tau)=-\frac{1}{{H}\tau}$, with $H =constant$ for the very early universe. In our previous works \cite{c35, c37}, an asymptotically dS solution in terms of index of the Hankel function $\nu$ ( with ${\nu}\neq{3/2}$) was considered as the fundamental mode during inflation. The best values of $\nu$ which are compatible with the scalar spectral index from the latest observational data \cite{c43} are $1.513\leq \nu \leq 1.519$ \cite{c44} or by best fitting with Planck data $( c\simeq1.05 )$ \cite{c45}. These values of $c$ motivated us to the departure from dS mode (30) to excited-dS modes (\ref{gen27}). Therefore, we use the asymptotic-dS modes as the fundamental initial modes for calculations in the next sections.

\section{Calculation with Asymptotic-dS Vacuum}
Because we do not know anything about the physics before inflation in the very early universe,
any primary excited vacuum mode can be considered as a good and acceptable mode for the initial state. Since the recent Planck results \cite{c43}, as observational evidence, motivated us to use a non-Bunch-Davies vacuum for quasi-dS slow roll inflation, let us as the logical choice consider an asymptotic-dS mode as the fundamental mode during quasi-dS inflation. The logic behind the choice of exited-dS mode is that the mode at early times ($\tau\rightarrow{-\infty}$) corresponds to an adiabatic mode, what one would naturally think of as the vacuum. For later times (when $\tau\rightarrow{0}$), the second and higher order terms of $\upsilon_{k}$ become important. We expect these additional terms to lead to particle creation, thereby providing the correction terms in the spectra.

According to (18), the general solution of the equation of motion in this quasi-dS space-time (${\nu}\approx\frac{3}{2}$), including positive and negative frequency, can be given by (\ref{gen27}),
  and we call it the \emph{excited-$\alpha$-vacuum}\cite{c38}. Note that the mode function (30) is the exact solution of Equation (4) for pure dS inflation, and similarly we consider mode (\ref{gen27}) as the general asymptotic solution of Equation (4) for quasi-dS inflation. Since this vacuum is of order $(\frac{1}{k\tau})^{2}$, for $g_{k}$ we again consider two different choices~\cite{c39,c46} of orders $1$ and $0$, i.e. $(\frac{1}{k\tau})^{1}$ and
$(\frac{1}{k\tau})^{0}$.\\

\subsection{Corrections to the \emph{dS Background}}
  If we consider quasi-dS space-time as the fundamental space-time during inflation with pure dS background, for the first choice of $g_{k}$, we can consider the $g^{(dS)}_{k}$ corresponding to the first derivative or conjugate momentum of $\upsilon_{k}$.  We will have
  \begin{equation} \label{g1}
g^{(dS)}_{k}=\sqrt\frac{k}{2}{A_{k}}(1-\frac{i}{k\tau})e^{-ik\tau}-\sqrt\frac{k}{2}{B_{k}}(1+\frac{i}{k\tau})e^{ik\tau}.
\end{equation}
We follow Section (2.3) in \cite{c1} and obtain for (\ref{gen27})
and (32), \begin{equation} \label{cofi11}
B_{k}=-\gamma_{k}A_{k}e^{-2ik\tau_{0}},\;\;
|A_{k}|^{2}=\frac{1}{1-|\gamma_{k}|^{2}},\end{equation} where \begin{equation}
\gamma_{k}=\frac{{i}(1-c)k\tau_{0}+d}{2(k\tau_{0})^{2}+i(c+1)k\tau_{0}-d}.\end{equation} If we ignore
the terms higher than second order, we obtain the corrected power
spectrum as a function of parameter $\sigma_{0}$,

$$P^{(dS)}_{\phi}(k)=(\frac{H}{\pi(1-\sqrt{8c+1})})^{2}\Big[\frac{1}{1-((c-1)\sigma_{0})^2}\Big]$$
\begin{equation} \label{pds}\times
\Big(1-(c-1)\sigma_{0}\sin(\frac{2}{\sigma_{0}})+
(\frac{(c-1)\sigma_{0}}{2})^{2}\Big).
\end{equation}
Since the background is selected as dS space-time, it is expected that correction terms for the pure dS case ($c =1$) in (\ref{pds}) disappear and we have a scale-invariant spectrum, but we expect to have scale-dependent spectra with non-zero correction terms for the any quasi-dS case with $c \neq 1$.

\subsection{Corrections to the \emph{flat background}}
On the other hand, if we consider quasi-dS space-time as the fundamental space-time during inflation with a flat background, the vacuum is chosen by requiring that the mode functions $\upsilon_{k}$ reduce to the Minkowski ones in the limit $\tau\rightarrow-\infty$. So, as the second choice, we consider $g^{(Fl)}_{k}$ in flat space-time as follows,

 \begin{equation} \label{gen30}
g^{(Fl)}_{k}=\sqrt\frac{k}{2}{A_{k}}e^{-ik\tau}-\sqrt\frac{k}{2}{B_{k}}e^{ik\tau}, \end{equation}
and for (\ref{gen27}) and (36) we again obtain $B_{k}$ and $A_{k}$
 similar to (33), but for $\gamma_{k}$ we have
 \begin{equation} \gamma_{k}=\frac{i{c}k\tau_{0}+d}{2k^{2}\tau_{0}^{2}+i{c}k\tau_{0}-d},\end{equation}
and we obtain
 \begin{equation}
|f_{k}|^{2}\sim\frac{1}{1-(c\sigma_{0})^{2}}\Big(1-c\sigma_{0}\sin(\frac{2}{\sigma_{0}})+
(\frac{c\sigma_{0}}{2})^{2}\Big).
\end{equation}
If we use the following Taylor expansion for $x=\sigma^2_{0}<1$,

\begin{equation} \sum x^{n}=\frac{1}{1-x}, \end{equation}
the higher order trans-Planckian corrections for $P^{(Fl)}$ are obtained as

$$ y=(\frac{2\pi}{H})^{2}P^{(Fl)}_{\phi}=(\frac{2}{1-\sqrt{8c+1}})^{2} $$
\begin{equation} \label{fin30}\times\Big(1-c\sigma_{0}\sin(\frac{2}{\sigma_{0}})+
\frac{5}{4}(c\sigma_{0})^{2}-(c\sigma_{0})^{3}\sin(\frac{2}{\sigma_{0}})+...\Big).
\end{equation}

\begin{figure}[b]
  \centering

    \includegraphics[height=.4\linewidth]{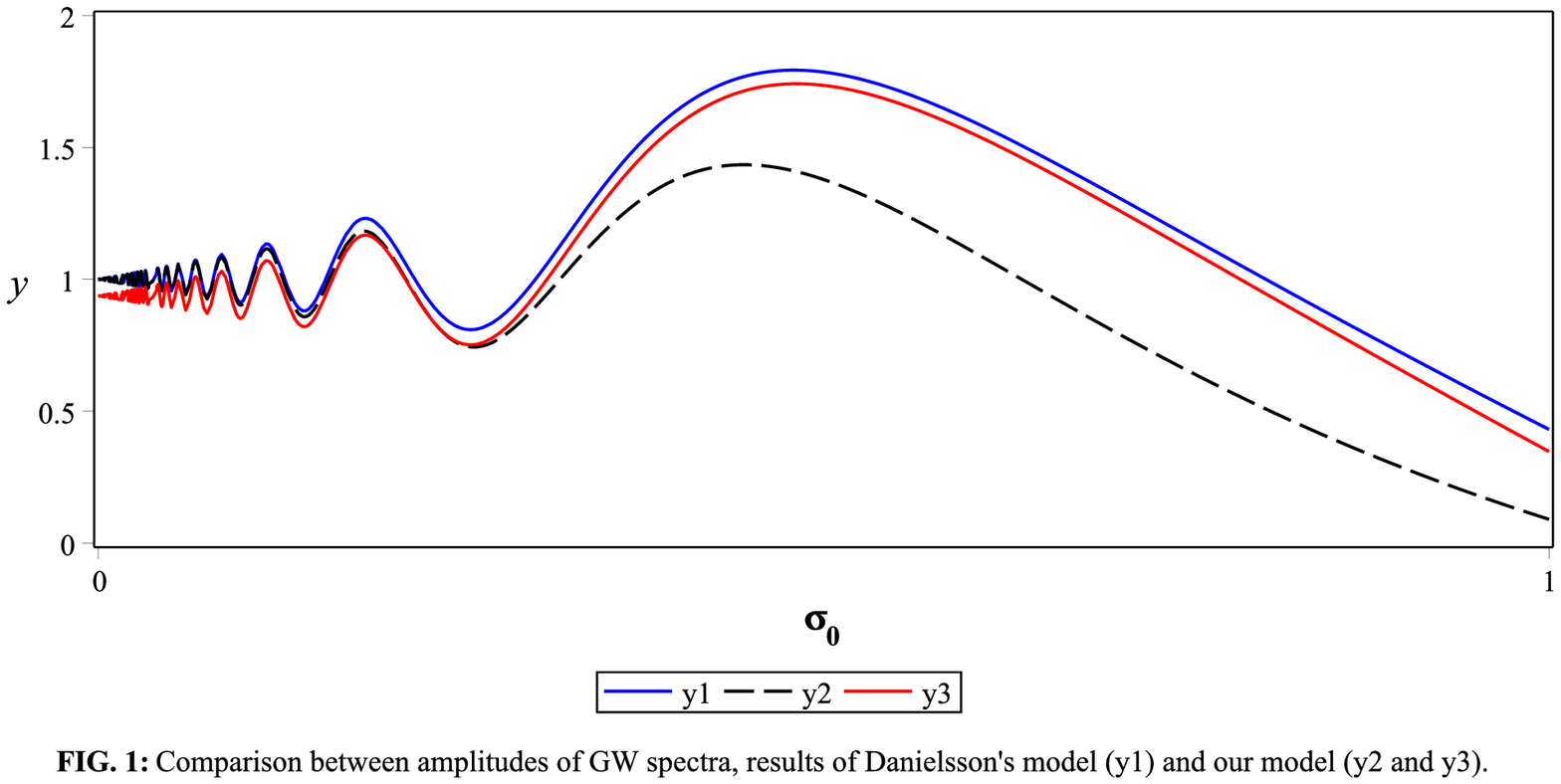}

      \end{figure}

As an important result of this method, the amplitude of the power spectrum is dependent on the choice of $c$, i.e. the type of space-time in the period of inflation. Also, the corrections in the final result (\ref{fin30}) includes first and higher order trans-Planckian corrections $ \sigma_{0}$. We see that from (\ref{fin30}), the corrections of spectra not only include higher order terms associated with the value of $c$ (type of  geometry), but also include the harmonic terms $(\sim\sin(\frac{2}{\sigma_{0}}))$, that may originate from the fluctuations of space-time (or gravitational waves) during inflation. Although these corrections are too tiny in the far past time limit, they may have very important effects on the CMB anisotropy and structure formation at the large scale. Also, our results in (\ref{fin30}) are theoretically valid for any asymptotic dS space-time and particularly confirm the conventional results for pure dS space-time \cite{c1} in first order. In Fig. 1., we see consistency of the power spectrum of our model with the power spectrum of Danielsson's model~\cite{c1}. In this figure, we plot the amplitude of the gravitational wave (GW) spectra of result $y_1$, and our result $y$ for particular values of $c$; $y_2$ for the dS case with $c =1$ and $y_3$ for the quasi-dS case with $c =1.05$. The latter value of $c$ is consistent with the data from Planck 2015 \cite{c43}. Also, in Fig. 2 , we compare the shift of the amplitude of the GW power spectrum originating from initial dS and quasi-dS modes for very small values of $\sigma_{0}$ that are theoretically important; for example, if $\Lambda$ is the Planck scale, $\sigma_{0}$ is at most $10^{-4}$ and if $\Lambda$ is the string scale, $\sigma_{0}$ could possibly be $10^{-2}$ \cite{c1}. As follows from these figures, for the trans-Planckian limit, the amplitude of the spectra is different from the Danielsson spectra and the shift of amplitude of the GW spectrum due to the initial quasi-dS mode $(c \neq{1})$ is more noticeable for $\sigma_{0}\ll{1}$ .\\

\begin{figure}[b]
  \centering

    \includegraphics[height=.4\linewidth]{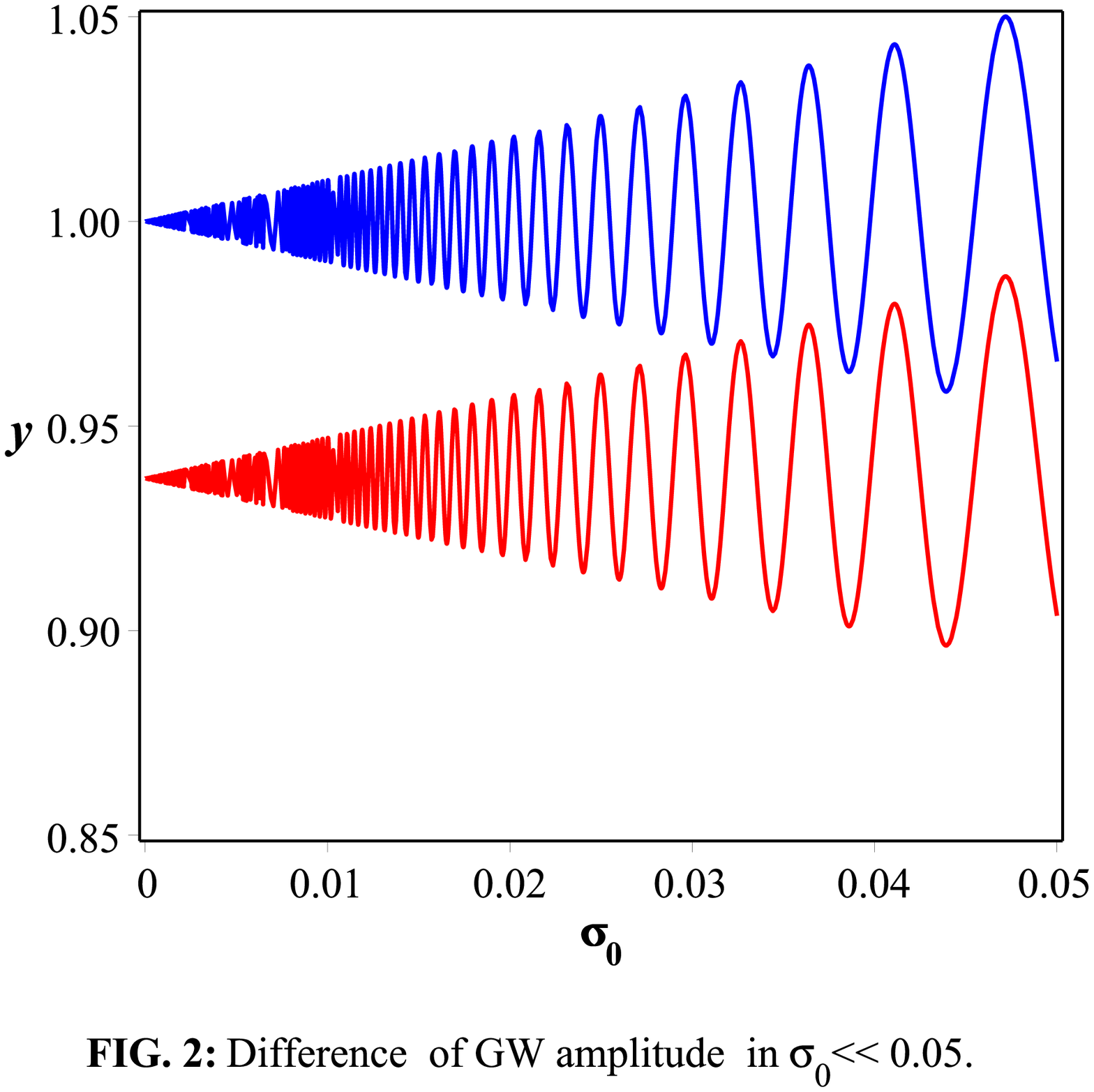}
    \includegraphics[height=.4\linewidth]{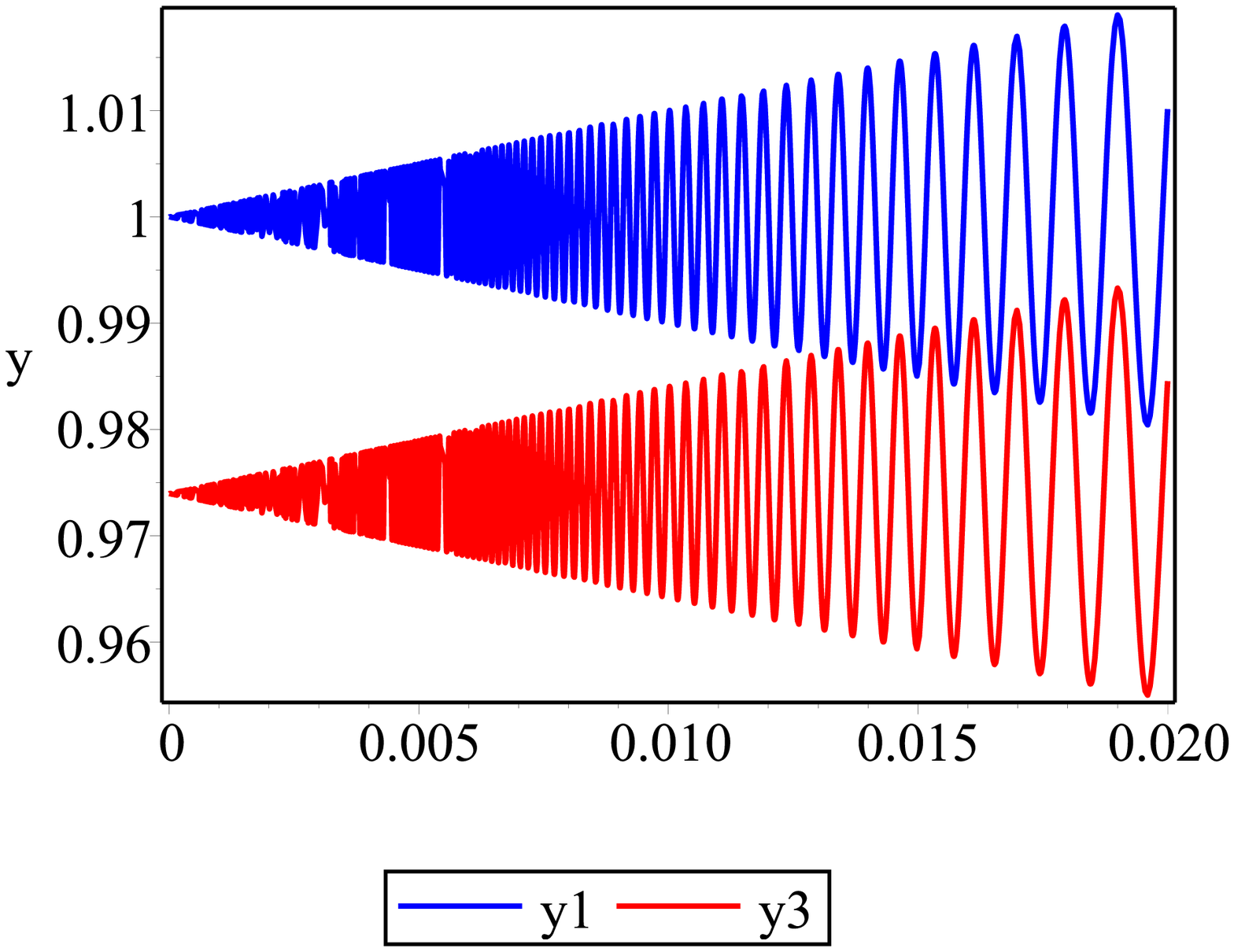}

      \end{figure}

In Ref.~\cite{c41}, the corrections are argued to be of size $ \sigma_{0}^2$, while in Refs.~\cite{c47, c48}, one is dealing with substantially larger corrections of order $ \sigma_{0}$. Also, it was argued in Ref.~\cite{c41}, using a low energy effective field theory, that local physics imply that the effects cannot be larger than  $ \sigma_{0}^2$. This was criticized in Ref.~\cite{c49}, where it was pointed out that the trans-Planckian physics can effectively provide the low energy theory with an excited vacuum, of course circumventing the arguments of Ref.~\cite{c41}. Finally, Martin and Brandenberger  obtained the corrections of all orders of $\sigma_{0}$ \cite{c50}, such as we obtain in (\ref{fin30}).

\section{Conclusions}
As we saw, taking a non-trivial initial mode instead of the Bunch-Davies mode results in an oscillatory spectrum. Furthermore, if the initial mode is of higher order too, it will also lead to higher order trans-Planckian corrections. This alternative initial mode  is essentially obtained by expanding the Hankel function for the quantum mode in quasi-dS spacetime up to higher order of $ \frac{1}{k\tau} $, which corresponds to quantization of finite wavelength rather than the ultraviolet limit. This higher order initial mode is reasonable, since it is an asymptotic expansion of the general solution of the Mukhanov-Sasaki equation for a quasi-dS background at very early time. For this excited initial mode, the slight deviation of the exact solution leads to the corrections and scale dependency of the power spectrum. The magnitude of the corrections can be expressed by the dimensionless value $\sigma_{0}$.\\
  The corrections and their amplitude  depend on the type of  geometry (i.e. different values of $c$), on the value and order of $ \sigma_{0} $, and  are also sensitive to the harmonic terms $(\sim\sin(\frac{2}{\sigma_{0}}))$ during inflation. Because some of the correction terms have oscillatory forms, it is likely that they are caused by the fluctuations of space-time known as gravitational waves. Finally, in the early universe with quasi-dS space-time, the probability of inflation with non-minimal coupling with gravity is possible, and we plan to investigate this issue and reconstruction of the initial excited dS mode by the Planck constraint in future works.\\

\noindent {\bf{Acknowledgements}}: We would like to acknowledge A Sojasi, H Moshafi, M and B Khanpour for their help in improving the manuscript. This work has been supported by the Islamic Azad University, Qom Branch, Qom, Iran.

\end{document}